\documentclass[a4paper,aps,prd,10pt,preprintnumbers,showpacs,twocolumn,superscriptaddress,nofootinbib,amsmath,amssymb,gensymb,amsfonts]{revtex4-2}
\usepackage{graphicx}
\usepackage{cmap}
\usepackage[utf8]{inputenc}
\usepackage[T1]{fontenc}
\usepackage{nicefrac}
\usepackage{bm,amsmath,amssymb,amsfonts,gensymb,bbold}
\usepackage[colorlinks=true,linktocpage=true,linkcolor=blue,citecolor=blue,allcolors=blue]{hyperref}
\usepackage{amsmath}
\usepackage{makecell}
\usepackage{soul}
\def\imo{i}
\def\re#1{Re(#1)}
\def\im#1{Im(#1)}
\def\K{{\cal K}}
\def\Order#1{{\cal O}\left(#1\right)}

\newcommand{\eq}[1]{\begin{align} #1 \end{align}}
\begin{document}
\title{Quantum corrected black holes: testing the correspondence between grey-body factors and quasinormal modes}
\author{Milena Skvortsova}\email{milenas577@mail.ru}
\affiliation{Peoples’ Friendship University of Russia (RUDN University),
6 Miklukho-Maklaya Street, Moscow, 117198, Russia}

\begin{abstract}
Grey-body factors and quasinormal modes are two distinct characteristics of radiation near black holes, each associated with different boundary conditions. Nevertheless, a correspondence exists between them, which we use to calculate the grey-body factors of three recently constructed quantum-corrected black hole models. Our findings demonstrate that the grey-body factors are significantly influenced by the quantum corrections for some of the models under consideration, and the correspondence holds with reasonable accuracy across all three models. We confirm that the grey-body factors are less sensitive to the near-horizon corrections of the spacetime, because the grey-body factors are reproduced via the correspondence using only the fundamental mode and the first overtone. 
\end{abstract}
\maketitle
\section{Introduction}

Grey-body factors and quasinormal modes of black holes are two distinct characteristics that arise from different boundary conditions. Grey-body factors measure the fraction of total radiation flux that penetrates the potential barrier and reaches a distant observer. They are calculated under the assumption that both incoming and outgoing radiation exist between the event horizon and the peak of the potential barrier. These boundary conditions correspond to a scattering problem and are frequently used in the analysis of black hole Hawking radiation \cite{Hawking:1975vcx,Page:1976df,Page:1976ki,Kanti:2004nr} or superradiance phenomena \cite{Starobinsky:1973aij,Starobinskil:1974nkd,Bekenstein:1998nt,Torres:2016iee,Konoplya:2008hj,Hod:2012zza,East:2017ovw}. 

In contrast, quasinormal modes are complex proper oscillation frequencies of black holes that do not permit incoming radiation either from the event horizon or from infinity. These modes dominate the black hole’s response to external perturbations during the ringdown phase and are observed by gravitational wave interferometers \cite{LIGOScientific:2016aoc,LIGOScientific:2017vwq}.

Despite their differing boundary conditions, recent research suggests a potential link between quasinormal modes and grey-body factors in the high-frequency regime \cite{Oshita:2023cjz,Okabayashi:2024qbz}. Grey-body factors, in fact, appear to be more stable geometric characteristics than quasinormal modes when small static deformations of the geometry are imposed \cite{Rosato:2024arw,Oshita:2024fzf}. An exact correspondence between quasinormal modes and grey-body factors has been established in the high-frequency limit for spherically symmetric black holes \cite{Konoplya:2024lir} and extended to certain axially symmetric black holes \cite{Konoplya:2024vuj}. This correspondence is precise in the eikonal limit and is based on the WKB approximation. Consequently, in cases where the eikonal limit is not accurately described by the WKB method—such as in some theories with higher curvature corrections \cite{Konoplya:2020bxa,Konoplya:2017wot,Bolokhov:2023dxq}—the correspondence may fail or may only apply to those parts of the spectrum that are well approximated by the WKB method \cite{Konoplya:2022gjp}. However, even for cases that support this correspondence, the accuracy of the relationship beyond the eikonal limit remains an open question.

To test this correspondence, it is necessary to select a range of black hole models and compare the grey-body factors derived from precise quasinormal mode calculations with accurate grey-body factors obtained numerically or through other sufficiently precise methods. In this work, we employ the 6th-order WKB approach \cite{Schutz:1985km, Iyer:1986np, Konoplya:2003ii} as a reliable method, at least for gravitational perturbations beginning from $\ell=2$, as demonstrated in numerous studies on both quasinormal modes and grey-body factors \cite{Skvortsova:2023zmj,Skvortsova:2024wly,Skvortsova:2024eqi,Malik:2024tuf,Malik:2024sxv,Malik:2023bxc,Dubinsky:2024aeu,Dubinsky:2024hmn,Kokkotas:2010zd,Stashko:2024wuq,Bolokhov:2023bwm,Konoplya:2006ar,Kodama:2009bf,Cuyubamba:2016cug}.

To test the correspondence, we consider a quantum-corrected black hole for which quasinormal modes have been previously calculated. The literature on various models of quantum-corrected black holes, their quasinormal spectra, and grey-body factors is extensive (see, for example, \cite{Saleh:2016pke,Abdalla:2005hu,Liu:2020ola,Yang:2022btw,Gogoi:2022wyv,Konoplya:2017ymp,Daghigh:2020fmw,Cruz:2020emz,Konoplya:2023ahd,Bolokhov:2023bwm,Chen:2022ngd}). 

The specific black hole models we examine in this test were recently obtained in \cite{Zhang:2024khj,Lewandowski:2022zce} following the general Hamiltonian constraints approach \cite{Thiemann:2007zz,Ashtekar:2004eh}. This framework yields two qualitatively distinct solutions. The first solution addresses the long-standing issue of general covariance in spherically symmetric gravity, which arises when constructing a semiclassical model of black holes within canonical quantum gravity \cite{Zhang:2024khj}. In this context, two distinct black hole models are proposed, each depending on the choice of a quantum parameter. The second solution emerges within the framework of the quantum Oppenheimer-Snyder model in loop quantum cosmology, where the energy-momentum tensor contributes quantum corrections to the Schwarzschild spacetime \cite{Lewandowski:2022zce}.

While quasinormal modes for these black hole models have been examined in several recent studies \cite{Gong:2023ghh,Skvortsova:2024atk,Zinhailo:2024kbq,Zinhailo:2024kbq,Luo:2024dxl,Malik:2024nhy}, no similar analysis has been conducted for grey-body factors, except in \cite{Heidari:2024bkm}, where grey-body factors were estimated for certain test fields. However, that study did not include gravitational perturbations, and were calculated with the low order WKB method implying insufficient accuracy. Thus, our work not only tests the correspondence but also supplements the existing literature on grey-body factors of quantum-corrected black holes.

Using the precise values of quasinormal modes found in \cite{Konoplya:2024lch}, we calculate the grey-body factors of the gravitational field through the correspondence developed in \cite{Konoplya:2024lir} up to order $\ell^{-2}$. We then compare these results with the 6th-order WKB formula, which is known to be highly accurate for gravitational perturbations. Our findings demonstrate that the correspondence is a sufficiently precise and effective tool for determining the grey-body factors of black holes, maintaining an error below a few percents for the near extreme black holes and within one percent for moderate values of the quantum coupling.

The paper is organized as follows. In Sec. II, we briefly describe the three black hole models and the wave equation for axial gravitational perturbations. Sec. III presents the calculation of grey-body factors using the correspondence with quasinormal modes and test of the accuracy of this correspondence. Finally, in the Conclusions, we summarize and discuss our results.

\section{Black hole metrics and wave-like equations}\label{sec:wavelike}

The study in \cite{Zhang:2024khj}  addresses maintaining covariance in spherically symmetric vacuum gravity by introducing an arbitrary effective Hamiltonian constraint and a freely chosen function for constructing the effective metric. The authors establish conditions on a Dirac observable representing black hole mass, leading to two families of quantum-corrected Hamiltonian constraints. These constraints yield distinct quantum-corrected metrics, with classical constraints recovered when quantum parameters are set to zero. 
The corresponding black-hole metric is given by the general line element
\begin{equation}\label{metric}
ds^2=-f(r)dt^2+\frac{1}{g(r)}dr^2+r^2(d\theta^2+\sin^2\theta d\phi^2),
\end{equation}
where for the\textit{ first type black-hole model}, the metric functions are
\begin{equation}
\label{eq:metric_model_1}
f(r)=g(r)= \left(1- \frac{2 M}{r}\right)\left[1+ \frac{\xi ^2}{r^2}\left(1-\frac{2 M}{r}\right)\right],
\end{equation}

and $\xi $ is the quantum parameter, while $M$ is the ADM mass.

The metric functions of the \textit{second black-hole model} constructed within the same approach \cite{Zhang:2024khj}  have the form:
$$
\begin{array}{rcl}
\label{eq:metric_model_2}
f(r)&=&\displaystyle 1-\frac{2 M}{r},\\
g(r)&=&\displaystyle f(r) \left(1+\frac{\xi ^2}{r^2}f(r)\right).\\
\end{array}
$$

Finally,  the metric function for the \textit{third black-hole model} was obtained in \cite{Lewandowski_2023},
\eq{\label{eq:metric_3}
f(r)=g(r)=1-\frac{2M}{r}+\frac{M^2\xi}{r^4},
}
where $\xi$ is the Barbero-Immirzi parameter. In this model  the event horizon exists for $\xi\leq 27M^2/16$.  In this work we follow the units used in earlier publications for calculations of quasinormal modes. Thus in the first teo midels we suppose $M=1/2$, while in the third model we use units of mass $M=1$.

After separation of the variables the axial type of gravitational perturbations can be reduced to the Schrödinger wavelike form \cite{Kokkotas:1999bd,Berti:2009kk,Konoplya:2011qq}:
\begin{equation}\label{wave-equation}
\dfrac{d^2 \Psi}{dr_*^2}+(\omega^2-V(r))\Psi=0,
\end{equation}
where the ``tortoise coordinate'' $r_*$ is defined as follows:
\begin{equation}\label{tortoise}
dr_*\equiv\frac{dr}{\sqrt{f(r) g(r)}}.
\end{equation}

\begin{figure*}
\resizebox{\linewidth}{!}{\includegraphics{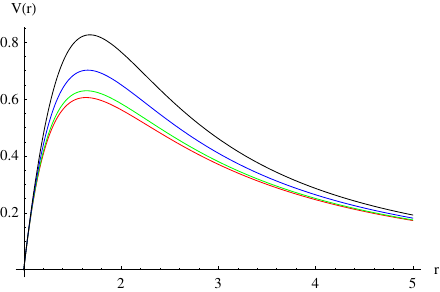}\includegraphics{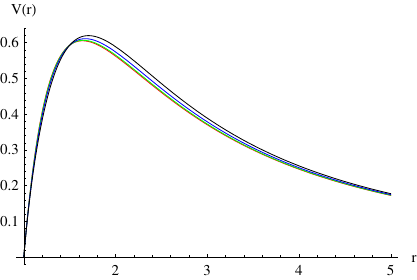}\includegraphics{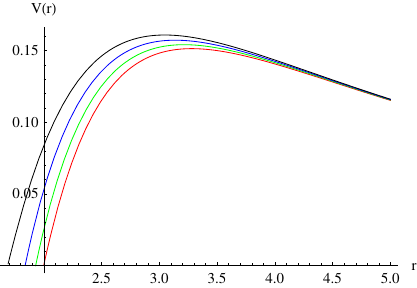}}
\caption{Effective potentials for various $\xi$ for the gravitational perturbations of the first (left), second (middle) and third (right) BH models. Here $M=1/2$ for the first and second models and $M=1$ for the third model,  $\xi=0$ (red), $\xi=0.5$ (green), $\xi=1$ (blue), $\xi=1.5$ (black).}\label{fig:potentials}
\end{figure*}

The effective potentials for axial gravitational perturbations have the form \cite{Konoplya:2024lch}
\begin{equation}\label{potentialScalar}
V(r)=f(r)\left(\frac{2g(r)}{r^2}-\frac{(fg)'}{2rf}+\frac{(\ell+2)(\ell-1)}{r^2}
\right).
\end{equation}
where $\ell=2, 3, 4, \ldots$ are the multipole numbers.
Examples of effective potentials for all three models are given in fig. \ref{fig:potentials}.  We see that the effective potentials are positive definite in the whole space from the event horizon to infinity. This means that the perturbations are stable and all the quasinormal modes must decay in time.  

\begin{figure*}
\resizebox{\linewidth}{!}{\includegraphics{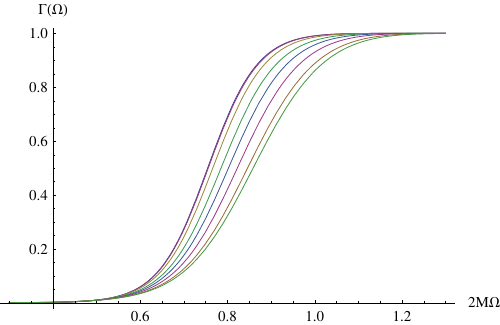}\includegraphics{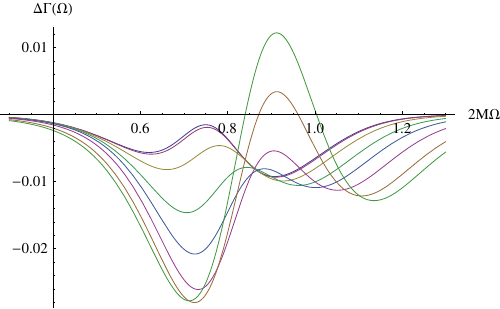}}
\caption{Right panel: Grey-body factors for various $\xi$ for the $\ell=2$ axial gravitational perturbations of the first BH model; the parameter $\xi =0.1$, $0.2$, $0.5$, $0.8$, $1$, $1.2$, $1.4$, $1.5$ from top to bottom, $M=1/2$. Left panel: difference between grey-body factors obtained by the 6th order WKB and the correspondence.}\label{fig:GBFmodel1}
\end{figure*}

\begin{figure*}
\resizebox{\linewidth}{!}{\includegraphics{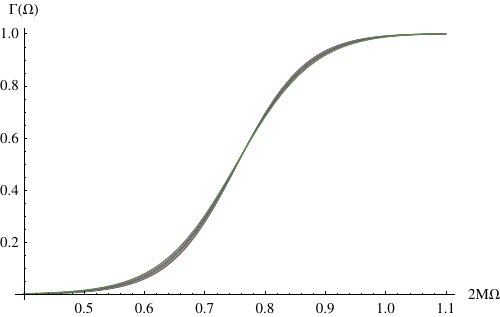}\includegraphics{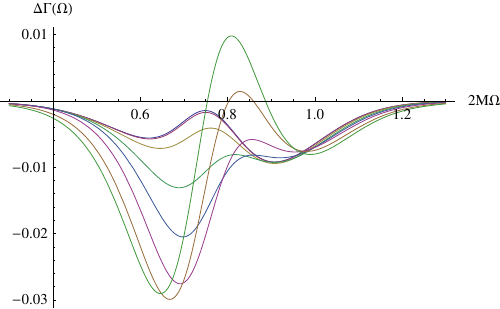}}
\caption{Right panel:Grey-body factors for various $\xi$ for the  $\ell=2$ axial gravitational perturbations of the second BH model; the parameter $\xi =0.1$, $0.2$, $0.5$, $0.8$, $1$, $1.2$, $1.4$, $1.5$, $M=1/2$. Left panel: difference between grey-body factors obtained by the 6th order WKB and the correspondence.}\label{fig:GBFmodel2}
\end{figure*}

\begin{figure*}
\resizebox{\linewidth}{!}{\includegraphics{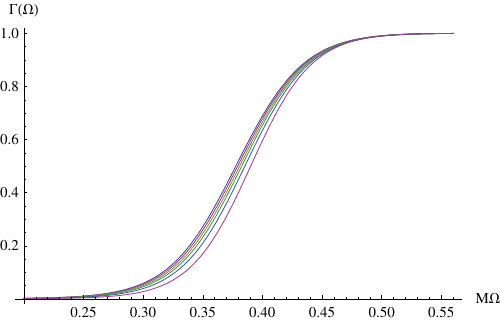}\includegraphics{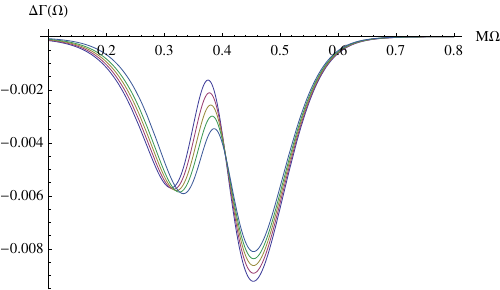}}
\caption{Right panel:Grey-body factors for various $\xi$ for the  $\ell=2$ axial gravitational perturbations of the third BH model; the parameter $\xi =0.1$, $0.3$, $0.5$, $0.7$, $1$, $1.5$, $M=1$. Left panel: difference between grey-body factors obtained by the 6th order WKB and the correspondence.}\label{fig:GBFmodel3}
\end{figure*}

\section{Grey-body factors obtained via the correspondence with quasinormal modes}

In scattering processes around black holes, partial reflection of a wave by the potential barrier produces the same grey-body factors whether the wave originates near the horizon or arrives from infinity. This symmetry leads to boundary conditions defined as follows:
\begin{equation}
\begin{array}{rclcl}
\Psi &=& e^{-i\Omega r_*} + R e^{i\Omega r_*}, &\quad& r_*\to+\infty, \\
\Psi &=& T e^{-i\Omega r_*}, &\quad& r_*\to-\infty,
\end{array}
\end{equation}
where $R$ and $T$ represent the reflection and transmission coefficients, respectively. In the context of black hole radiation, the transmission coefficient, $T$, is also referred to as the grey-body factor:
\begin{equation}
\Gamma_{\ell}(\Omega) =|T|^{2}=1-|R|^{2}
\end{equation}

The general form for the quasinormal frequencies in the WKB approximation can be expressed as a series expansion around the eikonal limit, as follows \cite{Konoplya:2019hlu}:
\begin{eqnarray}\label{WKBformula-spherical}
\omega^2 &=& V_0 + A_2(\K^2) + A_4(\K^2) + A_6(\K^2) + \ldots \\\nonumber
&-& i\K\sqrt{-2V_2}\left(1 + A_3(\K^2) + A_5(\K^2) + A_7(\K^2) + \ldots\right),
\end{eqnarray}
where $V_0$ is the potential at the peak, and $V_2$ is the second derivative of the potential with respect to the tortoise coordinate. Here, the terms $A_i(\K^2)$ denote higher-order WKB corrections. These corrections are explicitly detailed for the second and third WKB orders in \cite{Iyer:1986np}, for the fourth through sixth orders in \cite{Konoplya:2003ii}, and for orders up to the thirteenth in \cite{Matyjasek:2017psv}. In the scattering problem, the WKB corrections are applied similarly, though the expression for $\K$ differs in accordance with the distinct boundary conditions.

The WKB method has been extensively developed and utilized across a wide range of studies \cite{Konoplya:2001ji,Abdalla:2005hu,Paul:2023eep,Becar:2023zbl,Xia:2023zlf,Al-Badawi:2023lvx,Chen:2023akf}, making it a robust tool for analyzing both quasinormal modes and grey-body factors. Given the wealth of literature on this subject, we will not delve into further details of the WKB method here.

The correspondence between the grey-body factors and quasinormal modes were derived for the spherically symmetric and asymptotically flat black holes via the WKB expression for the grey-body factors,
\begin{equation}\label{eq:gbfactor}
\Gamma_{\ell}(\Omega)=\dfrac{1}{1+e^{2\pi\imo \K}},
\end{equation}
where \cite{Konoplya:2024lir}
\begin{eqnarray}\nonumber
&&\imo\K=\frac{\Omega^2-\re{\omega_0}^2}{4\re{\omega_0}\im{\omega_0}}\Biggl(1+\frac{(\re{\omega_0}-\re{\omega_1})^2}{32\im{\omega_0}^2}
\\\nonumber&&\qquad\qquad-\frac{3\im{\omega_0}-\im{\omega_1}}{24\im{\omega_0}}\Biggr)
-\frac{\re{\omega_0}-\re{\omega_1}}{16\im{\omega_0}}
\\\nonumber&& -\frac{(\omega^2-\re{\omega_0}^2)^2}{16\re{\omega_0}^3\im{\omega_0}}\left(1+\frac{\re{\omega_0}(\re{\omega_0}-\re{\omega_1})}{4\im{\omega_0}^2}\right)
\\\nonumber&& +\frac{(\omega^2-\re{\omega_0}^2)^3}{32\re{\omega_0}^5\im{\omega_0}}\Biggl(1+\frac{\re{\omega_0}(\re{\omega_0}-\re{\omega_1})}{4\im{\omega_0}^2}
\\\nonumber&&\qquad +\re{\omega_0}^2\Biggl(\frac{(\re{\omega_0}-\re{\omega_1})^2}{16\im{\omega_0}^4}
\\&&\qquad\qquad -\frac{3\im{\omega_0}-\im{\omega_1}}{12\im{\omega_0}}\Biggr)\Biggr)+ \Order{\frac{1}{\ell^3}}.
\label{eq:gbsecondorder}
\end{eqnarray}
Here $\omega_0$ and $\omega_1$ are, respectively, the dominant mode and the first overtone.

The grey-body factors are decreased when the quantum correction parameter $\xi$ is turned on (see figs. \ref{fig:GBFmodel1}-\ref{fig:GBFmodel3}). This can be explained by the behaviour of the effective potentials: when $\xi$ is increased, the potential barrier is increased and the transmission coefficient becomes smaller. This behaviour can be easily seen for the first and third black hole models. The effective potential for second model depends only slightly on $\xi$. Therefore, the grey-body factors are almost unaffected by the quantum correction for that model. 

An interesting example is given by the third black hole model for which the effective potential approaches the Schwarzschild one at a distance from the black hole, but drastically different in the near horizon zone. Overtones for this model experience an outburst, deviating from their Schwarzschild values at a speed increasing with the overtone number $n$. However, the grey-body factors are much less sensitive than overtones to the change of $\xi$, as seen in figs. \ref{fig:GBFmodel2}, because grey-body factors depend only on the fundamental mode and the first overtone which deviate from their Schwartzschild limits only moderately.

However, grey-body factors can be calculated with greater precision using the higher order WKB expression for $\K$. The grey-body factors have been found in this way for various black holes and wormholes 
\cite{Konoplya:2010kv,Fernando:2016ksb,Stashko:2024wuq,Dubinsky:2024nzo,Bolokhov:2024voa,Konoplya:2020jgt,Konoplya:2023ahd,Toshmatov:2015wga,Heidari:2023bww,Grain:2006dg,Dey:2018cws} and despite this is not exact expression for the quasinormal modes at finite $\ell$, we will use it to estimate the order of the relative error of the correspondence.

Comparison of the grey-body factors calculated via the 6th order WKB approach with those found through the correspondence with quasinormal modes shows that the difference ranges from a fraction of one percent to two-three percent depending on the value of $\xi$ and type of the black hole (see figs. \ref{fig:GBFmodel1}-\ref{fig:GBFmodel3}).  We have observed such a reasonable accuracy for the lowest multipole numbers $\ell =s$, where $s$ is the spin of the field. For larger $\ell$ the accuracy is even higher and usually do not exceed a small fraction of one percent. 

\vspace{5mm}

\section{Conclusions}

Recently a correspondence between grey-body factors and quasinormal modes has been established \cite{Konoplya:2024lir}. 
This correspondence is precise in the eikonal limit, but its 
accuracy is, strictly speaking,  unknown for finite $\ell$, due to asymptotic convergence of the WKB series. 
On the other hand, it was reported that unlike overtones of quasinormal modes, which are sensitive to small near horizon corrections \cite{Konoplya:2022pbc},  grey-body factors are much less sensitive to small corrections of the effective potential \cite{Rosato:2024arw,Oshita:2024fzf}.  Therefore testing the correspondence on metrics which deviate mainly in the near horizon zone  is also testing the superior stability of the grey-body factors. The best choice for such metrics are various quantum corrected black holes. We used three recent models of such black holes to test the correspondence and showed that it holds with the remarkable accuracy for all of them. We have also confirmed that relatively small near horizon deformations, unless they also essentially change the geometry near the peak of the effective potential, do not lead to considerable shift in the grey-body factors.

\begin{acknowledgments}
I would like to acknowledge R. A. Konoplya for sharing his numerical data of \cite{Konoplya:2024lch} and useful discussions. This work was supported by RUDN University research project FSSF-2023-0003.
\end{acknowledgments}

\bibliographystyle{unsrt}
\bibliography{bibliography}
\end{document}